

\input graphicx

\magnification=1200
\catcode `@=11

\font\tenll=lasy10
\newfam\llfam
\textfont\llfam=\tenll

\def\n{\noindent}

\def\b{\bigskip}
\def\m{\medskip}
\def\w{\widetilde}
\baselineskip 16pt
\centerline{\bf{Constrained effective potential in hot QCD}}
\bigskip

\centerline{by}
\bigskip
\centerline{C.P.Korthals Altes}
\bigskip
\centerline{Centre Physique Th\'eorique au CNRS}
\centerline{Campus de Luminy, Route L\'eon Lachamps, B.P.907}
\centerline{F13288, Marseille, Cedex 2. France}
\bigskip
\centerline{\bf{ABSTRACT}}
Constrained effective potentials in hot gauge theory give the probability
that a configuration p of the order parameter (Polyakov loop) occurs.
They are important in the analysis of surface effects and bubble formation in
the plasma. The vector
potential appears
non-linearly in the loop; in weak coupling the linear term gives rise
to the traditional
free energy graphs. But the non-linear terms generate insertions of the
constrained modes into the free energy graphs, through renormalisations of the
Polyakov loop. These insertions are gauge dependent and are necessary to cancel
the gauge dependence of the free energy graphs. The latter is shown, through
the BRST identities, to have again the form of constrained mode insertions.
It also follows, that absolute
minima of the potential are at the centergroup values of the loop.  We evaluate
the two-loop contributions for SU(N) gauge theories, with and without quarks,
for the full domain of the N-1 variables.

\vfill\eject

\noindent {\bf 1. Introduction}
\b
Study of high-temperature QCD$^{1)}$ has become urgent because of the building
of the relativistic ion collider (RHIC).
Apart from this, there is the vast area of early cosmology, where hot QCD has
come into play.
 At these temperatures the Z(N) symmetry of quarkless QCD is broken
spontaneously leaving us with N ordered phases. Adding quarks breaks this
symmetry explicitely, but the
deconfinement transition is still there.

Recently there has been interest$^{2)3)}$ in the surface tension that occurs in
QCD with quarks only as an order-disorder interface, and, subsequently, in the
surface tension between the different ordered phases in quarkless QCD. These
ordered phases are characterized by the order parameter (the Polyakov loop)
having a $Z(N)$ center group value. Numerical work on this preceded our
analytic approach$^{4)5)}$, and both are in reasonable agreement.

One key ingredient in these calculations is the profile $p$ that the Polyakov
loop $p$ develops in between two $Z(N)$ phases. The other key ingredient is the
probability that a given profile occurs. This quantity is called a
''constrained'' effective potential.

In this paper we will calculate the constrained effective potential $U$, that
is associated with the probability that a given constant mode $\bar p\equiv
{1\over V}\int d\bar x\ p(\bar x)$ appears :
$$\exp[-V\beta U(t)]\equiv{\int DA\delta(t-\bar p)\exp -{1\over g^2}S(A)\over
\int DA\exp -{1\over g^2}S(A)}$$

\noindent $U(t)$ as a function of $t$ is computable in perturbation theory. It
is accessible to Montecarlo simulations and is gauge independent. We will give
the perturbative rules for computing it ; these rules include the ad-hoc
prescription of ref 12.

In a previous paper$^{5)}$ we did compute the constrained effective action for
a special one parameter set of the Polyakov loop, in the absence of quarks.
This special set is called the ''$q$ valley''. Its physical significance is
due to the fact that it vehicles the tunneling effects giving rise to the
surface tension between two ordered vacua$^{5)}$.

In this paper we will give the full result, including all values of the
Polyakov
loop outside the q valley, and including quarks. We show  (in section 3)  that the ad hoc insertion method in $^{12})$ and $^{5})$  follows from starting from the constrained path integral.  Our results are
useful
 for calculations of
transition probabilities between various stable and unstable minima that appear
in the quark case, e.g. for the case of three colours and six quarks$^{15)}$.

Apart from this we have been motivated by two reasons

\item{i)} To lift some of the uncertainties in the literature on the pure
gluon results. Those readers that are only interested in the end results can
consult eqns 4.2, 4.3 (pure glue) 5.2 and 5.4 (fermions). Our pure glue
results are in agreement with ref 5 eqn 5.1, 5.13 and 5.14.
\m
\item{ii)} To gain some insight into the gauge-cancellations using BRST identities generalized to a non-trivial Polyakov loop background, and the
accompanying stabilisation of the $Z(N)$ minima due to extra vertices induced by the background.  There is some overlap with
ref 13, but our methods are complementary. The gauge artifact cancellations are
shown in 2.35 and 3.18.

As mentioned, there has been related work by various authors, with different
  motivations$^{6)7)}$, and different outcomes. With our method we compute in
various gauge fixing choices and find the same result. This fortifies our
confidence in our results. Besides, we find that the $Z(N)$ vacua stay stable
to
two loop order ; if quarks are present the stable vacuum stays at where the
Polyakov loop takes the value one, where as metastable minima remain at
roughly the location of the non-trivial center group values of the loop (see
section 5 and 6).

In section 2 we explain in some detail the method of calculation with as main result  eq. 2.35. The
expectation value of the Polyakov loop to one loop order is computed in section
3,  resulting in eq. 3.17 and 3.18. In section 4 the gluonic 2-loop results are given in 4.2 and 4.3.. In section 5 the
fermionic results are analysed. Section 6 contains conclusions and outlook. A few appendices
elaborate on the results.

\vfill\eject

\n{\bf 2. The method to compute the effective action $U$.}
\b
This section is rather long so it is divided into six subsections.
Section 2.a defines the effective action, section 2.b gives the saddle
point method for the perturbative evaluation and section 2.c deals
with the one loop results. In section 2.d the two loop results are
explained;
it is here that the intricacies of the non linearities of the Polyakov
loop enter. How the final answer is gauge choice independent is
detailed in sections 2.e and 2.f.

\bigskip

\n {\bf 2.a.} Effective potential $U$.
\m
We will work in a finite box $V=L^3$ in a heat both of temperature
$T\equiv\beta^{-1}$ ; in the box we have an $SU(N)$ Yang Mills field, and we
work on the usual Euclidean formulation with periodic b.c.'s in the time
direction. This means that the unitary NxN matrix $P(A_0(\vec x))$ defined by
$$P(A_0(\vec x))\equiv {\cal{P}}(\exp i\int^{\beta}_0d\tau A_0(\vec
x\tau))\equiv\lim_{n\rightarrow \infty}\prod^n_{k=1}\exp i\Delta\tau_kA_0(\vec
x,\tau_k)\eqno 2.1$$ - $\cal{P}$ being the path ordering,
$\tau_{k+1}\equiv\tau_k+\Delta\tau,\ \Delta\tau\equiv\beta/n$ - is transformed
by a periodic gauge transformation $\Omega$ like $\Omega P(A_0)\Omega^{-1}$,
and
has therefore gauge invariant eigenvalues.

Another way of saying this that :
$$t_1(A_0(\vec x))\equiv{1\over N}\ Tr\ P(A_0(\vec x)),\dots ,t_{N-1}(A_0(\vec
x))\equiv{1\over N}\ Tr\ P^{N-1}(A_0(\vec x))\eqno 2.1(a)$$
are gauge invariant. For $Tr\ P(A_0(\vec x))$, the ''Polyakov loop'' we have a
straightforward interpretation once it is averaged in the Euclidean path
integral formalism :
$$\left<t_1(A_0(\vec x))\right>\equiv{\int DA t_1(A_0(\vec x))\exp -{1\over
g^2}S(A)\over\int
DA\exp  -{1\over g^2}S(A)}\eqno 2.2$$
This average is the exponential of the free energy of an infinitely heavy
quark (i.e. a fermion in the fundamental representation of $SU(N)$) at the
point $\vec x$. $\left<t_1(\vec x)\right>$ depends on the spatial boundary
conditions. This simple relation between the average and the
free energy $F_q$ of a single quark reads:
$$\left<t_1(A_0(\vec x))\right>=\exp -\beta F_q\eqno 2.3$$
{}From this relation $^{1)}$ follows the use of $<t_1(\vec x)>$ as an order
parameter.

Averages of traces of higher powers of the loop 2.1 $t_k(\vec x)$ can be gotten
by suitable linear combinations
of free energies of ''quarks'' in higher representations of SU(N).
Taking the SU(3) case as an example, one can relate
$$<TrP(A_0)^2>=<TrP_6(A_0)>-<Tr\bar P(A_0)>$$
with $P_6$ the Polyakov loop in the sextet representation of SU(3).
{}From now on we will be interested in spatial averages $\bar t_k\equiv{1\over
V}\int d\vec x\ t_k(A_0(\vec x))$, and we will define, following Fukuda et
al$^{8)9)}$,
the constrained effective potential $U$ by:
$$\exp -\beta V (U(t_1\dots t_{N-1})+F)\equiv\int
DA\prod^{N-1}_{k=1}\delta(t_k-\bar t_k)\exp -{1\over g^2}S(A)\eqno 2.4$$
The left hand side is up to normalisation the probability that a given set of
fixed  numerical  values $t_1\dots t_{N-1}$ for the corresponding spatial averages  appears in our system$^{20}$. The normalisation is in
terms of the free energy F, which one gets by doing the path integral
2.4 {\it without} the constraints. In ref 9 it was argued that this
effective potential is numerically a very useful quantity, and easier
accessible than say the usual free energy in the presence of sources $J_1\dots
J_{N-1}$ for the $\bar t_k$ :
$$\exp -\beta VF(J_1\dots J_{N-1})\equiv \int DA\exp -{1\over
g^2}S(A)+V\sum^{N-1}_{k=1} J_k\cdot\bar t_k$$
This free energy is obviously related to 2.4 through the Laplace transform :
$$\eqalignno{
\exp -\beta VF(J_1\dots J_{N-1})=&\int dt_1\dots
dt_{N-1}\exp{VJ_1t_1+\dots +VJ_{N-1}t_{N-1}}\cr
&\quad\exp - \beta VU(t_1\dots t_{N-1})&2.5\cr}$$
$U$ and $F$ are clearly gauge invariant.

 Consider the infinite
volume limit in eqn 2.5.  The ensuing saddle point equations give  $U$  the same dependence
on
the $t_k$ as the effective potential $G=F-{\partial F\over \partial J}.J$  has on
the expectation values of the $t_k$. In fact one has
$$G=U+F(0)\eqno 2.6$$
in the infinite volume limit.
The reader may ask at this point: why not discuss the effective
potential instead of the constrained effective potential?
A first reason for preferring the latter is that it is given directly in terms
of a
path integral, eqn 2.4. Second, in a finite volume there are
differences, which may become significant in perturbation theory, when
there are infrared divergencies. Third, the study of surface effects
necessitates the use of the constrained effective potential.

Some authors$^{13)}$ prefer to discuss the eigenvalues of the loop,
rather than the traces. The Jacobian between the two is a van der
Monde
determinant, as explained in appendix D. We prefer traces, because they
are easier to handle when taking higher order effects into account
(see section 3).

The next subsection deals with the perturbative evaluation of the effective
potential.

\bigskip

\n{\bf 2.b.} Perturbative evaluation.
\m
As in ref 5) we will Fourier transform the $\delta$-constraints. This will give
us $N-1$ variables $\lambda_k,\ k=1\dots N-1$, with :
$$\prod^{N-1}_{k=1}\delta(t_k-\bar t_k)=\int\prod^{N-1}_{k=1} d\lambda_k\
e^{i{\lambda_k\over g^2}(t_k-\bar t_k))}\eqno 2.7$$
In the following we will drop the index $k$ for notational simplicity,
but indicate the dependence of $\bar t$ on the potential $A_0$.
Substituting 2.7 into 2.4 leaves us with a path integral in terms at the gauge
invariant variables $\lambda$ and the gauge potentials $A$ :
$$\exp -\beta VU(t)=\int d\lambda\int DA\exp i\lambda(t-\bar t(A_0))-{1\over
g^2}S(A)\eqno 2.8$$

To keep the bookkeeping of our degrees of freedom precise, we have to
introduce boundary conditions, e.g. periodic boundary conditions.

We will search for a saddlepoint $\lambda=b,\ A=B$ and set therefore :
$$\eqalignno{
A_{\mu}&=B_{\mu}+gQ_{\mu}\cr
\lambda&=b+gq&2.9\cr}$$
where $Q$ and $q$ are quantumvariables, expand the exponent in 2.8 and require
that linear terms cancel.
The integration over the quantum variables in 2.9 will leave us with a
functional
$Z(b,B)$  which stays invariant under a gauge
transformation
$$A^{\Omega}=\Omega A\Omega^{-1}+\Omega\partial\Omega^{-1}$$
The latter can be rewritten in obvious way as
$$B^{\Omega}=\Omega B\Omega^{-1}+\Omega\partial\Omega^{-1}\eqno 2.10$$
together with
$$Q^{\Omega}=\Omega Q\Omega^{-1}$$
It is then clear that indeed
$$Z(b,B)=Z(b,B^{\Omega})\eqno 2.11$$
The path integral 2.8 still needs gauge fixing ; in order to retain
property 2.11, we will take the familiar
background field gauge fixing :
$$S_{g.f.}\equiv{1\over 2\xi}Tr(D_{\mu}(B)Q_{\mu})^2\eqno 2.12$$
with the Faddeev-Popov term :
$$\bar{\eta}D_{\mu}(B)D_{\mu}(B+Q)\omega\eqno 2.13$$

Using 2.1 and 2.9 we find for the linear term in an obvious short hand notation
:
$$i(t-\bar t(B))q+ib\bar t^{(1)}(B)Q-S^{(1)}(B)\cdot Q=0\eqno 2.14$$
This gives three saddle point equations, one for the $q's$ (2.15(a)), one for
those $Q's$, that couple to $\bar t^{(1)}(B)$ \underbar{and} $S^{(1)}(B)$
(2.15(b)) and finally those $Q's$, that couple \underbar{only} to
$S^{(1)}(B)$~:
$$\eqalignno{
t-\bar t(B)&=0&2.15(a)\cr
ib\bar t^{(1)}(B)-S^{(1)}(B)&=0&2.15(b)\cr
S^{(1)}(B)&=0&2.15(c)\cr}$$

Since 2.11 determines $B$ up to a gauge transformation we choose a solution $B$
that is computationally most convenient :
$$B_{\mu}=C\delta_{\mu0}\eqno 2.16$$
$C$ is a space time independent diagonal matrix in the Lie algebra of $SU(N)$ ;
this satisfies 2.15 with $b=0$. We emphasize this is just a choice, and will
come back to it in section 2.e, where the gauge variance of our results is
discussed.

So the eigenvalues $C_i (i=1\dots N)$ of $C$ obey a constraint
$$\sum^N_{i=1}C_i=0\eqno 2.17$$
and are related to the $N-1\quad t_k\quad(k=1,\dots ,N-1)$
 through $t_k={1\over N}Tr \exp ikC\beta$.

The advantage of this choice for the saddle point lies in the easy
diagonalisation of the quadratic part
$S^{(2)}$ of the action :
$$\eqalignno{
S^{(2)}=&iq{1\over V}\int d\vec xd\tau Tr\bar
t^{(1)}(C)Q_0(\vec x\tau)\cr
&+\int d\vec xd\tau\
TrQ_{\mu}(-D^2(C)+(1-\xi)D_{\mu}(C)D_v(C))Q_{\nu}+\cr
&2\int d\vec
xd\tau Tr\bar{\eta}(-D^2(C))\omega&2.18\cr}$$
$D_{\mu}(C)$ is the covariant
derivative : $D_{\mu}(C)\equiv\partial_{\mu}+\delta_{\mu 0}[C,$.

Let us diagonalize the last two terms in 2.18, in terms of the colour basis
$(\lambda^{ij})_{n,m}={1\over\sqrt{2}}\delta_{in},\delta_{jm}$ and
$\lambda^d\equiv{1\over r_d}diag(1,\dots,1,1-d,0\dots,0),\ Tr\ \lambda_d=0$.
Details on this  Cartan basis are discussed in appendix D.

The Fourier transform of the $Q's$ is defined as :
$$ Q_{\mu}(p)\equiv{1\over V}\int d\vec x\int^{\beta}_0d\tau
e^{-i\vec p\vec x-ip^0\tau}Q_{\mu}(\vec x,\tau)\eqno 2.19$$
with
$$\left.\matrix{
p^0&\equiv&2\pi Tn_0\cr
\vec p&\equiv&{2\pi\over L}\vec n\cr}\right\}|n_{\mu}|=0,1,2,\dots\eqno
2.19(a)$$

We write $Q_{\mu}(p)$ in terms of colour components as :
$$Q_{\mu}(p)=\sum_{i\not
=j}Q^{ij}_{\mu}(p)\lambda^{ij}+\sum^N_{d=2}Q^d_{\mu}(p)\lambda^d\eqno 2.20$$
and find for $S^{(2)}$ in its diagonalized form, restoring the colour indices:

$$\eqalignno{
S^{(2)}&=i\sum^{N-1}_{l=1,d}q_l\bar t^{(1)}_{l,d}(C) Q^d_0(0)+{V\over
2}T\sum_{i\not =j,n_{\mu}} Q^{ij}_{\mu}(p)((p^{ij})^2\delta_{\mu
v}-(1-\xi)p^{ij}_{\mu}p^{ij}_v) Q^{ji}_v(-p)\cr
&+{V\over 2}T\sum^N_{d=2}\sum_n Q^d_{\mu}(p)(p^2\delta_{\mu
v}-(1-\xi)p_{\mu}p_v) Q^d_v(-p)\cr
&+VT\sum_{i\not =j}\sum_{n_{\mu}}{\bar{\eta}}^{ij}
(p)(p^{ij})^2{\omega}^{ij}(-p)\cr
&+VT\sum_{d=2}^N\sum_{n_{\mu}}{\bar{\eta}}^d
(p)p^2{\omega}^d(-p)&2.21\cr}$$

In the first term the matrix $\bar t^{(1)}_{l,d}=Tr\exp(ilC)\lambda_d$ results from expanding the 
l-th power of the Polyakov loop to first order in the quantum field $Q_0^d(0)$.
The symbol $p^{ij}_{\mu}$ stands for $p^{ij}_0=p_0+C_i-C_j$, and
$p^{ij}_k=p_k$. This shift in the momentum is due to the covariant derivatives
in 2.18. 

 Let us note that ${\bar{\eta}}^k(p=0),\w{\omega}^k(p=0)$ and
$\w Q^k_{\mu}(p=0)$ are non-Gaussian variables. This is due to our
periodic boundary conditions. We can use e.g. twisted b.c!s
and overcome this problem to get the same thermodynamic limit as we will get
here by ignoring the problem.

The $N-1$ constraint variables $q$ couple only to the $N-1$ zero-momentum
diagonal $Q_0$ variables, $ Q^d_0(0)$, in the first term of 2.21.

\bigskip

\n{\bf 2.c.} One loop result

The one loop result is obtained from the expression for $S^{(2)}$ in 2.21, by
substituting it in the exponent of 2.8.

We have first to integrate out the $q$ variables. This will give us delta
function constraints on the N-1 $Q^d_0(0)$ variables, of the type
$\delta(Q^d_0(0))$, and the one loop result is:
$$V\beta U^{(1)}(t)={1\over 2}Tr'\log(p^2_{ij}\delta_{\mu
v}-(1-\xi)p^{ij}_{\mu}p^{ij}_v)-Tr'\log p^2_{ij}\eqno 2.22$$

The prime on the trace means we left out the eigenvalues corresponding to these
$N-1$ modes.

We are interested in the thermodynamic limit $V\rightarrow\infty$. Then these
$N-1$ constraints, as argued in ref 5), are not important.

 For completeness we mention the result$^{1)}$ :
$$U^{(1)}(t)={2T^4\pi^2\over 3}\sum_{i \not =j}C^2_{ij}(1-C_{ij})^2\eqno 2.23$$
where $C_{ij}\equiv{C_i-C_j\over 2\pi T}$,and have to be taken mod 1; the $t$
are given in terms of
the diagonal matrix $C$ 2.16 :
$$t_k\equiv{1\over N}Tre^{ikC\beta}\eqno 2.24$$
Note that $U^{(1)}$ is independent of the gauge choice, as it should.
So far, all our labours did not lead to anything new. This changes in the next
subsection.

\b

\n{\bf 2.d.} Two loop results
\m
For the two loop results we need to expand the action and the constraint to
order $g^2$. To avoid clutter in the formulas we will stop indicating the
colour degrees of freedom on $q$, $\bar t^{(2)}(C)$ etc\dots. This leads to :
$$\eqalignno{ S^{(3)}=&iq({g\over 2!}\bar t^{(2)}(C)\cdot\overline{
Q^2_0}+{g^2\over 3!}\bar t^{(3)}(C)\cdot\overline{Q^3_0})\cr
&+{g\over 3!}S^{(3)}_{inv}\cdot Q^3+{g^2\over 4!}S^{(4)}_{inv}\cdot
Q^4+2(D_{\mu}(C)\bar{\eta})[Q_{\mu}\omega]&2.25\cr}$$
The bars mean the
average over the space-volume $V$.

$S^{(3)}$ is standard, except for the first term linear in $q$. Had the
constraint been linear in the potential, the second and third derivative of
$\bar t(C)$ would have been absent. Then we would have had - as in the one-loop
case - as only effect from the $q$-integration the $\delta$-function constraint
on the $N-1$ $Q^d_0(p=0)$ variables ; hence, in the thermodynamic limit, only
the traditional two-loop contributions with propagators and vertices carrying
momenta $p^{ij}_{\mu}$ instead of $p_{\mu}$, as depicted in the three graphs in
fig 1.

However, the linear terms in $q$ in eqn 2.25 {\it do} contribute through a
zero-momentum insertion, using the identity :
$$\eqalignno{
\int dQ\int dqe^{-iq\bar t^{(1)}Q}iq&=\int{dQ\over \bar
t^{(1)}}{-\partial\over\partial Q}\left(\int dqe^{-iq\bar t^{(1)}Q}\right)\cr
&=\int{dQ\over \bar t^{(1)^2}}\delta(Q){\partial\over\partial Q}&2.26\cr}$$
We have suppressed the $C$ dependence in $\bar t^{(1)}$ to simplify notation,
and $Q\equiv Q_0(0)$.This identity is of course nothing but the
expansion of the $\delta$ function in the defining equation
2.4.\footnote{$^*$}{The last step in this identity, the integration by parts,
is strictly only permitted if the boundary conditions are such that the $Q$'s
are Gaussian variables.} It gives us the zero-momentum insertions
through the derivative.

The identity 2.26 carries no volume factors, since the $\bar t$ has none right
from the outset, see eqn 2.4 and above. Moreover, when it acts on a given term
with a product of $Q$'s, only terms linear in $Q^d_0(0)$ will survive the
$\delta$-function in 2.26. If there are $n$ powers of $q$, then 2.26 changes to
one where $n$ derivatives of $Q$ act to the right.

After these remarks we go through the usual procedure of expanding $S^{(3)}$
out of the exponential in 2.8 and doing the path integral, by contracting
$Q$-fields. Apart from the terms, that give rise to the diagrams in fig 1, we
have from the derivative in eqn 2.26 acting on $S^{(3)}$ and $(S^{(3)})^2$ :
$$\eqalignno{
\exp -\beta VU=&\int D'QD\bar{\eta}D\omega\exp\left[-Q\cdot S^{(2)}\cdot
Q-\bar{\eta}\cdot D^2(C)\cdot \omega\right]\cr
&g^2\left[{1\over 2!}{1\over \left\Vert\bar
t^{(1)}(C)\right\Vert^2}\left\{{1\over 2!}\bar
t^{(3)}(C)\sum_{p}Q_0(p)Q_0(-p)\right.\right.\cr
&-2\bar t^{(2)}(C)\sum_{p}Q_0(p)Q_0(-p)&2.27\cr
&\left({1\over 2!}S^{(3)}_{inv\atop
\mu v0}(C)\sum_{p}Q_{\mu}(p)Q_v(-p)\right.\cr
&\left.\left.\vphantom{{1\over 2!}\bar t^{(3)}(C)\sum_{P}}
+\sum_{p}D_0(C)\bar{\eta}(p)\omega(-p)\right)
\right\}\cr
&\left.+{1\over 2!}{1\over\Vert\bar t^{(1)}(C)\Vert^3}
\left(\bar t^{(2)}(C)\right)^2\sum_{p}Q_0(p)Q_0(-p)\right]\cr}$$
$\Vert\bar t^{(1)}(C)\Vert$ is the determinant of the matrix
$\bar t^{(1)}(C)$ in the first term of eqn 2.21.This determinant is
the van der Monde determinant formed from the matrix C in eqn 2.16, as
shown
at the end of appendix D.

Without knowledge of the explicit colour structure of eqn 2.27 the
reader can easily check the following.
All but one of the contractions in 2.27 are one loop contractions and hence of
order $0(1)$. The {\it only} one that is $O(V)$ comes from the two-loop
contraction we get from the second term in the straight brackets ; we have presented this
term pictorially in fig 2 and denote it by $U^{(2)}_p$. The reader can easily
verify this, using B.2 for the propagator.

This contraction consists of two factors ; one is give by the renormalisation
of
the Polyakov loop :
$$g^2\bar t^{(2)}(C)\sum_p\left<Q_0(p)Q_0(-p)\right>\eqno 2.28(a)$$
The other is given by the zero-momentum insertion and equals :
$${\partial\over\partial C}U^{(1)}(C)\eqno 2.28(b)$$
with $U^{(1)}(C)$ given in 2.23. Hence $U^{(2)}_p$ is the product of 2.28(a)
and (b).

Thus to two loop order $U^{(2)}$ consists of two parts :
$$U^{(2)}=U^{(2)}_f+U^{(2)}_P\eqno 2.29$$
$U^{(2)}_f$ is given by the graphs in fig 1 with the topology of the graphs
contributing to the free energy but with the energies shifted through the
background $C$, as described underneath 2.21.
$U^{(2)}_P$ is given by the 1-loop result 2.23 with the argument $C$ shifted by
the renormalisation of the loop 2.28(a)(see fig.2).

$U^{(2)}_f$ is known to depend$^{6)}$ on the gauge choice $\xi$.
$U^{(2)}_P$ depends on the gauge choice only through 2.28(a), since 2.28(b) is
independent of $\xi$ (see 2.23). Therefore there must be a relation between the
gauge artifacts in both terms, in order to cancel and give a $\xi$-independent
$U^{(2)}$.

This relation is provided in the next subsection.

\b

\n{\bf 2.e.} BRST and gauge variation.
\m
In this subsection we will establish a relation between the BRST prediction
for the gauge - variance of $U^{(2)}_f$ and the zero-momentum insertion 2.28
(fig 2(a)). In fact they are the same as a very simple argument will show.

In fig 1 we have the three diagrams that contribute $U^{(2)}_f$ to
$U^{(2)}$. In the $\xi$
gauge there are contributions linear, quadratic and cubic in $(1-\xi)$ from the
propagators. The latter two are absent, as shown in appendix A. The terms
linear in $(1-\xi)$ do not cancel. Rather, a simple observation shows they come
in with a factor 3 from graph 1a, 2 from 1b and 1 from 1c. Combining this with
the combinatorial factors in front we see in fig 3(a) that $U^{(2)}_f$ has
a term linear in $(1-\xi)$ of the form~:
$$-{(1-\xi)\over 2}T\sum_{p_0}\int{d^{n-1}p\over(2\pi)^{n-1}}\left(\sum_{i\not =
j}{p^{ij}_{\mu}\Pi^{(ij)}_{\mu v}p^{ij}_v\over
(p^{ij})^4}+\sum^N_{d=2}{p_{\mu}\Pi^{(d)}_{\mu v}p_v\over p^4}\right)\eqno
2.30$$
where $\Pi_{\mu v}$ is the one-loop self energy,  and the
momenta $p^{ij}$ have colour shifted Matsubara frequencies  defined underneath 2.21.

The first sum is over the off-diagonal propagators, the second over the
diagonal ones (in colour space).

In the zero temperature case we obviously would obtain zero for the
coefficient, because the BRST identities$^{10)}$ tell us that the self energies
are transverse.

However, at non-zero temperature the BRST identities are not constrained by
Lorentz (Euclidean) invariance. Although the
BRST identities are local and therefore still valid at finite temperature their
consequences are different. In our case, with a background field, they take the
form$^{11)}$ :
$$0={\delta\Gamma\over\delta Q}\cdot{\delta\Gamma\over \delta J}+\hbox{ghost
contribution + eventual matter contributions}$$
$\Gamma$ is the one-particle irreducible generating functional in an external
field $C$ ; at $Q=0$ it is related to $U_f$ by $\Gamma=\beta VU_f$, since $C$
is constant in space time. $J$ is the external source coupling to the gauge
variation of $Q$, the quantum field, through the term :
$$J\cdot D(Q+C)\omega\eqno
2.31$$
The dot means summation over all degrees of freedom.

To obtain information on the self energy $\Pi$ in eqn 2.30 one derives
the BRST identity with
 respect to the
ghost-field $\omega$ and the quantum field $Q$ ; setting all fields and sources
to zero we get :
$$0={\delta^2\Gamma^{(1)}\over\delta Q\delta Q}\cdot{\delta^2\Gamma^{(0)}\over
\delta J\delta\omega}+
{\delta\Gamma^{(1)}\over\delta Q}\cdot{\delta^3\Gamma^{(0)}\over
\delta J\delta\omega\delta Q}\eqno 2.32$$

This is the result to one-loop order. The first term in eqn 2.32
contains a double derivative with respect to the quantum fields. This
double derivative equals the self energy $\Pi$.
  All other terms are identically zero to
this order. At zero temperature Lorentz invariance renders the last term in eqn
2.32 identically zero, and the one-loop self energy is transverse. However at
finite temperature ${\delta\Gamma^{(1)}\over \delta Q_0(z)}$ is \underbar{not}
vanishing, and - by inspection- it equals the zero
momentum
insertion $\beta V{\delta U^{(1)}\over\delta C}$ (see fig 3(c)) ;
precisely through this zero-momentum insertion the renormalisation effect of
the Polyakov loop did couple to the effective action, eqn 2.28.

Note that only the temporal component survives in the second term of 2.32.

So we have that both the gauge artifacts of the graphs in fig 1 and the full
renormalisation of the Polyakov loop couple to one and the same thing : the
zero-momentum insertion of $Q_0$ into the 1-loop result.

In order to make contact with the next section on the renormalisation of the
Polyakov-loop we rewrite 2.32 in momentum space, with momenta and indices of the Cartan basis explicit :
$$0=ip^{ij}_{\mu}\Pi^{ij}_{\mu\nu}-i{\delta U^{(1)}\over \delta
Q^d_0(0)}f^{ij,ji,d}\delta_{\nu,0}\qquad 1\leq i\not = j\leq N\eqno 2.33(a)$$
and
$$0=ip_{\mu}\Pi^d_{\mu \nu}\qquad d=2,\dots,N\eqno 2.33(b)$$
The colour-shifted momenta $p^{ij}$ are defined below 2.21. $Q_0^d(0)$ is the Euclidean
space-time average of $Q^d_0(x)$
To obtain eqn 2.33(b) we used the fact that ${\delta\Gamma^{(1)}\over \delta
Q_0^{ij}}$ vanishes identically because of colour conservation. So only the off
diagonal self-energies are non-transverse. 

The terms linear in the
gauge parameter $(1-\xi)$ in $U^{(2)}_f$  are 
 proportional to the zero-momentum insertion in 2.33(a):
$${\delta U^{(1)}(C)\over {\delta{Q_0^d(0)}}}=\sum_{k,l}\widehat B_3(C_{kl})f^{kl,lk,d}.\eqno 2.34$$ 

When combining 2.33(a) and 2.34~ we use the identity\footnote{$^{*)}$}{
As follows by inserting the definition  D.2 for the structure constants and using the trace properties of the $\lambda^{ij}$.} :
$$\sum_{d=2}^Nf^{kl,lk,d}f^{ij,ji,d}={1\over 2}(\delta_{ik}+\delta_{jl}-\delta_{il}-\delta_{jk}).$$
So finally substituting 2.33(a) into 2.30 we obtain for the coefficient of $(1-\xi)$ in $U_f^{(2)}$ :
$$U_f^{(2)}=U_f^{(2)}(\xi=1)-2(1-\xi)g^2N\sum_{i<j}\widehat B_1(C_{ij})\widehat B_3(C_{ij}).\eqno 2.35$$

Both in 2.34 and 2.35 we used eq. C.5 (with d=4 and k=1, 2 respectively)  from Appendix C.

The $\xi=1$ contribution in 2.35 is trivial to obtain and will be  discussed in section 4.

Let us note that the presence of fermions does not change the form of the BRST
identity 2.32. So when fermions are present gauge artifacts can be treated the
same way. The fermionic contribution  shows up additively in the $\widehat B_3$ factor in 2.35, as is clear from fig. 3 (c).

\b

\n{\bf 2.f} Comparison of gauge variation of $U^{(2)}_f$ and $U^{(2)}_P$

\m
In this section we have found that both the renormalisation of the Polyakov loop, and the gauge
artifacts in the free energy part of the effective action are coupled to the same
object : the temporal and colour diagonal zero-momentum insertions into the one
loop result 2.23 (see 2.28 (b) and the combination of 2.30 and 2.33(a)). The gauge variations of $U^{(2)}_P$ and $U^{(2)}_f$ have the same structure. For
the former we
combine 2.28(a) and (b), for the latter 2.30(a) and 2.33(a). In section 3, eq. 3.18, we
will see that evaluation of the renormalisation of the loop (2.28(a)) actually
matches  the gauge artifacts in 2.35, but with opposite sign.

Therefore we will have complete cancellation of the
gauge artifacts in the sum of the two, in two loop order. One expects this to 
continue for any order.

\vfill\eject

\n{\bf 3. Renormalisation of the Polyakov-loop}
\b
The renormalisation of the loop is \underbar{not} the renormalisation of the
numerical value $<\bar t_1>\equiv {1\over V}\int d\vec x <t_1(\vec x)>$.

Rather, what renormalizes is the \underbar{relation} between $<\bar t_1>$ and
the
particular choice of saddlepoint $C$ that we made in eqn 2.16 in the phase of the loop. This
renormalisation can - and will - contain a reflection of our gauge choice 2.12.

As a general expectation one would say that the phase of the loop renormalizes without any
extra ultraviolet infinities, beyond the usual ones that renormalise couplings
and fermion masses. This should be so, because all these effects are thermal in
nature.

In fact it turns out, that the expectation values of the
gauge invariant traces $\bar t_1,\bar t_2,\dots\bar t_{N-1}$
indeed stay finite through one loop.   From them one can reconstruct the finite renormalizations of the eigenvalues of the loop easily.    As a caveat let us look how proceeding through the renormalization of the gauge variant unitary matrix $P(A_0)$  leads to a correction with unwanted properties.. 

So take the special unitary
diagonal matrix $P(A_0)=e^{i(C+gQ_0}$,.
 It
becomes after adding the one loop correction in fig 2.a a diagonal matrix
which is \underbar{not} special, \underbar{not} unitary and \underbar{not}
finite ! Nethertheless the expectation value of its trace $\bar t_1$ is finite as it should. The same
holds true for the quantum average of the $k^{th}$ power of the Polyakov loop
and taking its trace $\bar t_k$ (see appendix B).


We have for the average of the loop matrix to order $g^2$ :
$$\eqalignno{
{1\over V}\int d\bar
x<P(A_0)>&=e^{iC\beta}-g^2\int^{\beta}_0d\tau_1\int^{\tau_1}_0d\tau_2e^
{iC\tau_2}
\cr
&\left<Q_0(\tau_2)e^{iC(\tau_1-\tau_2)}Q_0(\tau_1)e^{iC(\beta-\tau_1)}\right>+
0(g^4)
&3.1\cr}$$
where we used the definition 2.1 for $P(A_0)$, 2.9 and 2.16. There is no $\vec
x$ dependence after contraction of the $Q_0$ fields in 3.1, so we drop any
reference to it for notational convenience.

The order $g^2$ correction receives a contribution from all diagonal quantum
fields $Q^d_0$ equal to :
$$-g^2\int^{\beta}_0d\tau_1\int^{\tau_1}_0d\tau_2e^{iC\beta}
\left<Q_0^d(\tau_2)Q_0^d(\tau_1)\right>\eqno 3.2$$
because the diagonal $Q^d_0$ do commute with the diagonal $e^{iC\tau_i}$. Since
\vfill\eject
$$\eqalignno{
\Delta_{00}(\tau_2-\tau_1)&\equiv
\left<Q_0^d(\tau_2)Q_0^{d}(\tau_1)\right>\cr
&=T\sum_{p_0}\int {d\vec p\over (2\pi)^3}\left({1\over p^2_0+\vec
p^2}-(1-\xi){p^2_0\over(p^2_0+\vec p^2)^2}\right)e^{ip_0(\tau_2-\tau_1)}&
3.3\cr}$$
we find with dimensional regularisation for the sum of all diagonal
contributions to 3.2:
$$-g^2{1\over 2}{N-1\over N}\left\{{\beta^2\over 2}T\int{d^{n-1}\bar p\over
(2\pi)^{n-1}}\left[{1\over\vec p^2}+2T\sum_{p_0\not =0}{1\over i
p_0}\Delta_{00}(p_0,\bar p)\right]\right\}\eqno 3.4$$
The first term is zero ; the second as well, since $\Delta_{00}$ is even in
$p_0$.

So we are left with the off-diagonal contributions
$Q^{ij}_0(\tau_2)\lambda^{ij}(i\not = j)$. Let us denote the
product of two $\lambda$'s by :
$$\lambda^{ij}\lambda^{ji}\equiv D^{ii}\quad i\not = j\ \hbox{fixed}\eqno 3.5$$
and
$$\lambda^{ji}\lambda^{ij}\equiv D^{jj}\quad i\not = j\ \hbox{fixed}\eqno 3.6$$
$D^{ii}$ is the diagonal matrix, with all entries zero except the $i^{th}$
diagonal  element, which equals 1/2.

We use the identity
$$e^{iC\tau}\lambda^{ij}e^{-iC\tau}=e^{i(C_i-C_j)\tau}\lambda^{ij}\eqno 3.7$$
and the propagator for the $Q^{ij}_0$ excitation :
$$\Delta^{ij}_{00}\equiv\left<Q_0^{ij}(\tau_2)Q_0^{ji}(\tau_1)\right>
=T\sum_{p_0}\int{d\vec p\over(2\pi)^3}
\left\{{1\over
(p_0^{ij})^2+\vec p^2}-(1-\xi){(p^{ij}_0)^2\over((p^{ij}_0)^2+\vec
p^2)^2}\right\}e^{ip_0(\tau_2-\tau_1)}\eqno 3.8$$
Notice the occurrence of $p^{ij}_0\equiv p_0+C_i-C_j$. The time dependence of
the propagator is of course periodic.

With the help of 3.5 to 3.8 we can rewrite the $O(g^2)$ contribution in the
form (see appendix B for details) :
$$\eqalignno{
g^2\sum_{1\leq i\leq j\leq
N}&\left[\left\{\left(\vphantom{{\beta\over
i}}e^{-i(C_i-C_j)\beta}-1\right)D^{ii}+\left(e^{i(C_i-C_j)
\beta}-1\right)D^{jj}\right\}\Delta^{ij}_{(2)}\right.\cr
&\left.-{\beta\over
i}(D^{ii}-D^{jj})\Delta^{ij}_{(1)}\right]e^{iC\beta}&3.9\cr}$$
with
$$\Delta_{(r)}^{ij}\equiv T\sum_{p_0}\int {d^{n-1}\bar
p\over(2\pi)^{n-1}}{1\over (p_0^{ij})^r}\Delta_{00}(p^{ij}_0,\bar p)\eqno
3.10$$
The terms proportional  $D^{ii}$, $D^{jj}$ come from $i<j\ ,\
i>j$ respectively.

The factors ${1\over (p_0^{ij})^r}$ in 3.10 stem from the integrations over the
$\tau$-variables : they are there because of the non-locality of the loop.
Observe in 3.9 that the term with r=1 has a similar momentum structure
as  in 2.30, i.e. the loop on the left in fig 3.c proportional to $\widehat B_1$ as in 2.35. We will see below, that
the part with r=2 is projected out, after taking the trace of eq. (3.9).

Obviously the renormalisation of the loop in 3.9 gives again a diagonal matrix and      so eqn 3.1 can be rewritten as
$${1\over V}\int d\vec x <P(A_0)>=\exp i(C+g^2\delta C)\beta+0(g^4)\eqno 3.11$$
with $\delta C$ given by the matrix
$$\eqalignno{
{1\over 2}\sum_{1\leq i\not = j\leq
N}&\left[{T\over
i}\left\{\left(e^{-i(C_i-C_j)\beta}-1\right)D^{ii}+\left(e^{i(C_i-C_j)
\beta}-1\right)D^{jj}\right\}\Delta^{ij}_{(2)}\right.\cr
&\left.\vphantom{{T\over
i}}+(D^{ii}-D^{jj})\Delta^{ij}_{(1)}\right]&3.12\cr}$$
We note that the trace of $\delta C$ is \underbar{not} zero ! It is :
$$Tr\delta C={T\over 2i}\sum_{1\leq i\not = j\leq
N}(cos(C_i-C_j)\beta-1)\Delta^{ij}_{(2)}\eqno 3.13$$
Note that $\Delta^{ij}_{(2)}$ is logarithmically divergent whereas
$\Delta^{ij}_{(1)}$ is finite (see eqns C.5, C.15 and C.17). So our caveat
at the
beginning of this section has come true : as a matrix the expectation value
of
 the Polyakov loop has
become infinite, non-unitary and not special.

One might argue, that the Polyakov loop as a matrix is not gauge
invariant; in particular under a periodic and constant (in $\vec x$)
gauge transformations 3.11 is similarity transformed, and so is the
exponent
$C+g^2\delta C$. But the trace 3.13 stays invariant, so the infinity
cannot be absorbed by the gauge transformation!

On the contrary, when we first calculate the expectation value of the
{\it traces} of 
 the  $N-1$ powers of the loop, $\bar t_1,\bar t_2\dots\bar
t_{N-1}$
 , there is
{\it no} infinity, and the traces are {\it real}.
To see this, let us start with $SU(2)$ $(N=2)$. Then computing the trace of 3.9
gives us
$$<\bar t_1>=\cos \beta(C_1-C_2)-g^2{1\over 2}\beta\Delta^{12}_{(1)}\sin
(C_1-C_2)\beta\eqno 3.14(a)$$
with
$$\Delta^{ij}_{(1)}={1\over 4\pi}(2+(1-\xi))\left(C_{ij}-{1\over
2}\right)\eqno
3.14(b)$$
The trace projects out the infinite terms ! The result 3.14 coincides
with that of Belyaev$^{12)}$ and that of ref 5.

For $SU(N)$ we find for the $<\bar t_k>$ (see appendix B) :
$$\eqalignno{
<\bar t_k>=&{1\over N}Tr\left<(P(A_0))^k\right>={1\over N}Tr\
e^{ikC}\cr
&-{g^2\over N}{\beta\over i}{k\over
2}\sum_{i,j}Tr(D^{ii}-D^{jj})e^{ikC}\Delta^{(ij)}_{(1)}\qquad k=1,\dots
,N-1&3.15\cr}$$
All infinite contributions do cancel in the trace. These $N-1$ results
in
eqn 3.15 can be summarized by adding to $C$ a
diagonal traceless matrix $\underline{\delta}C$ (not to be confused with the
matrix $\delta C$ in 3.11 and 3.12), and to demand that the $k$-th power of
this matrix gives us the trace computed in 3.15 :
$$<\bar t_k>={1\over N}Tr\exp
ik(C+\underline{\delta}C)\beta+C(g^4)\eqno 3.16$$ Identifying 3.16 with 3.15
gives for a diagonal element : $${\underline{\delta}C_i\over 2\pi T}={g^2\over
(4\pi)^2}(2+1-\xi)\sum_jB_1(C_{ij})\eqno 3.17$$
where $B_1(x)\equiv x-{1\over 2}\epsilon(x)$. (See
appendix C).

{}From the anti-symmetry of $\Delta^{ij}_{(1)}$ in its argument
$C_{ij}$, it follows that the loop does not renormalize when $C=0$, as expected. This is
useful for knowledge of absolute minima (see section 6). It also shows
that $\sum_{i=1}^N\underline{\delta} C_i=0$, i.e. the matrix $\underline{
\delta} C$ is indeed traceless (and real and finite).

To find the full result for the Polyakov loop inserted  $U^{(2)}_P$ we combine 2.28(a) and (b)
with 3.17 to get:
$$ U^{(2)}_P= 2(2+(1-\xi))g^2N\sum_{i<j}\widehat B_1(C_{ij})\widehat B_3(C_{ij})\eqno 3.18$$

Eqns 3.17 and 3.18 are the main results of this section\footnote{$^*$}{To obtain 3.18 more in line with the derivation of 2.35, take the insertion   from 2.28  as $\sum_k< \bar t_k ><{\partial\over{\partial {(\bar t ^{(1)}}{k_,d}Q_0^d(0))}}S_{int}>$. By inserting into  3.15 a complete set of diagonal generators $\lambda_d$  one retrieves  the matrix $\bar t^{(1)}_{l,k}$ defined below 2.21. It factors out  and therefore cancels the same matrix in the denominator of the average. Use of 2.34 and the subsequent identity
for the f-symbols yields 3.18.} . They generalize the
results in ref. 5 beyond the $q$-valley. They will be used when evaluating the
full constrained effective action in the next two sections.

\vfill\eject

\n{\bf 4. The gluonic constrained effective potential}
\b
In this section we will give the one and two loop results. Everything is
expressed in terms of the Bernouilli polynomials $B_k(x)$, that we have
compiled in appendix C.

The result for an $SU(N)$ gauge theory becomes to two loop order :
$$U=U^{(1)}+U^{(2)}\eqno 4.1$$
with
$$U^{(1)}={4\pi^2\over 3}T^4\sum_{i<j}\tilde B_4(C_{ij})\eqno 4.2$$
and from 2.35 and 3.18:
$$
U^{(2)}=U^{(2)}_f(\xi=1)+U^{(2)}_P(\xi=1),\eqno 4.3$$
\noindent with the righthand side given in eq. (4.7) and (4.8). As anticipated
all gauge artifacts have dropped out.
The variable $C_{ij}$ equals
$$C_{ij}\equiv{C_i-C_j\over 2\pi T}\eqno 4.4$$
as in eqn 2.23, and $\tilde B_4$, $\tilde B_2$ are defined in eqns C.13
and C.14.


A few remarks. $U^{(1)}$ and $U^{(2)}$ are both
of order $N^2$ for a generic value of $C_{ij}$. However, for the specific
choice in ref 5
: $C_1=C_2=\dots=C_{N-1}={q\over N}$ and $C_N=-{(N-1)\over N}q$ we have only
$N-1$ terms : $$U^{(1)}={4\pi^2\over 3}T^4(N-1)\tilde B_4(q)\eqno 4.5$$
$$U^{(2)}={4\pi^2\over 3}T^4\cdot -5{g^2N\over (4\pi)^2}(N-1)\tilde B_4(q)\eqno
4.6$$

So in this one parameter space the potentiel is $O(N)$ smaller, and we will
denote it by the "$q$-valley".

The reader can see this in the $SU(3)$ example in fig 4.

The $q$-valley formulae 4.5 and 4.6 were derived in ref 5 (eqn 5.14 in ref 5).
Remarkable is the simplicity of 4.6: it just renormalizes
multiplicatively the one loop result 4.5. This is lost outside of the
q-valley.



The actual derivation of $U^{(2)}_f$ in  4.3 is astonishingly simple in $\xi=1$ gauge.
First we compute the contribution from the 3 graphs in fig 1. Using momentum
conservation like in A.3, all integrands are of the form ${1\over l^2_il^2_j}$,
where $l_i\ ,\ l_j$ are two different momenta from the set of momenta $l_1\
l_2$ and $l_3$ flowing through the lines of fig 1a and b. The answer is :
$$U_f=g^2\sum_{b,c,a}|f^{bca}|^2(\widehat B_2(C_c)\widehat B_2(C_a)-\widehat
B^2_2(0))\eqno 4.7$$
The colour indices $b,c,a$ run through the basis $(ij)$ and $d$.
$C_c={C_i-C_j\over 2\pi T}$ if $c=(i,j)$, $C_c=0$ if $c=d$ ($d$ labelling a
diagonal generator).

The contribution from the Polyakov loop  inserion $(\xi=1)$ is from 3.18 ~:
$$U_P^{(2)}=4g^2N\left[\sum_{i<j}\widehat B_3(C_{ij})\widehat
B_1(C_{ij})\right]\eqno 4.8$$
where the $\widehat B_k$ are momentun integrals defined in appendix C
and up to some multiplicative factors identical to the Bernoulli
polynomials
used before.
Now the total $U_f+U_P$ reduces in the q-valley to 4.6, using the formulae in appendix C, and
properties of the structure constants in appendix D.

We would like to draw the readers attention to a remarkable, but unwanted property
of $U_f$ in 4.7. Let us write out its content in the q-valley:
$$U^{(2)}_f={g^2N\over{4}}.{(N-1)\over 3}T^4[3q^2(1-q)^2-2q(1-q)]\eqno 4.9$$
We see that the linear term causes an absolute minimum, where the
value of $U^{(2)}_f$ is negative. But remember the discussion in
section 2.d: the free energy graphs alone follow from the
$\delta$
function constraint linear in the potential. As 4.7 is normalised by
the
the free energy graphs at C=0 we would expect the constrained path
integral
to give us a {\it non-negative} result! The fact, that it is not, just
illustrates once more how gauge dependent effective potentials can
give deceptive information. As another example$^{5)}$ take this
absolute
minimum $q_m$ of U to two loop order. It will be of order $O(g^2)$. But by
charge
conjugation we get $-q_m$ and it is not hard to see, that there we
have another absolute minimum: charge conjugation seems spontaneously
broken!

Both unwanted properties do disappear when we add 4.8, to get 4.6. More
generally, also outside the q-valley our result 4.3  is non-negative
and has only Z(N) minima.
\vfill\eject

\n{\bf 5. Fermions}
\b
In this section we add $n_f$ quarks (taken to be massless) in the fundamental
representation of $SU(N)$. As fermions they are \underbar{anti}-periodic in
time
$\tau$. The action reads
$$S=\int d\vec xd\tau\bar q(\not\!{\partial}+\not\!\!{A})q\eqno 5.1$$
for every flavour. The Z(N) invariance is now broken through the
boundary conditions: a gauge transformation which is periodic modulo a
Z(N) phase will change the anti periodicity.

 We go through the same arguments as in section 2.b, so
introduce background and fluctuation variables for the gauge fields.
The next step is to compute the contribution $U^{(1)}_q$ of the quarks to the
constrained effective potential ; this has been done$^{1)}$ and gives :
$$U^{(1)}_q(t)=-{4\over 3}\pi^2T^4n_f\left(\sum_i\tilde B_4\left({C_i\over 2\pi
T}+{1\over
2}\right)-{N\over 16}\right)\eqno 5.2$$
for $n_f$ quark flavours. The minus sign in front of 5.2 is due to the
fermionic determinant, the Bernoulli function appears just as in the
gluon
case 4.2. Its argument is shifted over ${1\over 2}$ because of the
anti-periodicity. The result is normalised to zero for $C_i=0$.
In what follows we will use the abbreviation :
$$C^f_i\equiv {C_i\over 2\pi T}+{1\over 2}\eqno 5.3$$
To two loop order we will find-like we did for the pure gluon
case in section 2.d - that the renormalisation of the Polyakov loop couples to
the zero-momentum insertion of $U^{(1)}_q$. Likewise the arguments in section
2.e are valid for the linear gauge variation of the fermion case (see
figs1, 2, and 3).

The two loop result for $n_f$ quarks becomes :
$$\eqalignno{
U^{(2)}_q=-n_f4\pi^2T^4{g^2\over(4\pi)^2}&\left.\left[\sum_{i\not
=j}\left\{\vphantom{{8\over
3}\sum_j}B_2(C_{ij})(B_2(C^f_i)+B_2(C^f_j))-B_2(C^f_i)B_2(C^f_j)\right\}\right.
\right.\cr
&+{N-1\over N}\sum_i(2B_2(0)B_2(C^f_i)-B^2_2(C^f_i))\cr &\left.-{8\over
3}\sum_iB_3(C^f_i)\sum_jB_1(C_{ij})+(N^2-1){5\over 144}\right]&5.4\cr}$$

The terms in the braces come from the configuration in fig 1(d) where the gluon
is off-diagonal. The next term stems from the diagonal gluons (see eqn D.7).
The one but last term in 5.4 comes from the zero momentum insertion into 5.2  using eqn
3.17, and restores the minimum at $C_i=0$. Thus, also in the fermion
case
charge conjugation and non-negativity are restored!
But the Z(N) minima are no longer degenerate, because the symmetry is
explicitely broken.

In the $q$-valley {\it both} gluon contribution and fermion contribution are
of order $N$, when $q\simeq N/2$. This explains why the fermion contribution
is comparable to the gluon contribution (fig 6).

For $SU(3)$ we plotted in
fig 4 and 5 the full potential 4.2, 4.3 and 5.2, 5.4 with $n_f=0, 2$
respectively. The
''$q$-valley'' is seen on the border of the admitted values of the loop. It is
plotted for $n_f=6$ in fig 6.

\vfill\eject

\n{\bf 6. Conclusion and outlook}
\b
In this paper we established the perturbative expansion of the constrained
effective action, to two-loop order in the large volume limit. Our result is
consistent with the
following prescription, to all orders :

\item{i)} Compute the sum $U_f(C)$ of all 1-PI free energy diagrams, with the
energies shifted by $C$, and obtain
$$\w U(C)\equiv U_f(C)-U_f(0)\eqno 6.1$$
\m
\item{ii)} Compute the renormalisation of $C$ to all orders to obtain
$$C=C(C_r)\eqno 6.2$$.
\m
\item{iii)}
$$\w U(C(C_r))\equiv U(C_r)\eqno 6.3$$
is the final answer.

Of course the result 6.3 is intuitively expected because of the
relation
2.6 between U and the effective potential G

As a function of $C_r\ ,\ U$ will not depend on the gauge choice. $U$ is
therefore perfectly well defined in contrast to statements made by some
authors. Of course, it may be that in three and higher loop order
infrared divergencies will show up. They fall into two categories. The
first one concerns divergencies due to the diagonal gluon
propagators. The off diagonal propagators are protected by a value of
$C_{ij}$ of order 1. The second category corresponds to small values
of $C_{ij}$ of order $g$. The first category can be compared to the
infrared
divergencies in an Abelian $U(1)^{N-1}$ theory and is less severe.
Some inroads into the latter have been made in ref 6).
The
second one is truly non Abelian and more severe.

The other important issue was that the effective potential is
non-negative
and has Z(N) minima as absolute minima. A discussion of the latter can
be found in ref. 13). Here we want to point out that the two are related.
To this end, suppose
$$U(C_r)\geq 0\eqno 6.4$$
to all orders, because, at least formally, $\exp -\beta VU(C_r)$ is a
probability.
We would like to know wether $U(0)$ is an absolute minimum, i.e. wether
$U(0)=0$. According to 6.1, 6.2 and 6.3:

$$U(C_r=0)=0\eqno 6.5$$
is equivalent to :
$$F(C(C_r=0))=F(0)\eqno 6.6$$
The simplest way to have 6.6 fulfilled is
$$C(C_r=0)=0\eqno 6.7$$
This is actually the case to two loop order (eqn 3.17).

The authors of ref (13) have found a very ingenuous gauge choice, which they
call Static Background Gauge (SBG) : $$S_{g.f.}={1\over 2\xi}\left({1\over
\xi'}D_0(C)Q_0+D_iQ_i\right)^2\eqno 6.8$$
In this gauge $\xi$ is fixed,
$\xi'\rightarrow 0$. It looks very probable that in this gauge the Polyakov
loop
does not renormalize to any order of perturbation theory. If so, then indeed
6.5 is true to any order in perturbation theory..

With our gauge choice we found in section 3, that only
traces of powers of loops are finite. The infinities
in 3.12 are {\it not} ultra violet in nature ; they stem from the factors
${1\over(p^{ij}_0)^2}$ that originate in the $\tau$-integrations in 3.1, that
is, in the non-locality of the loop. It has consequences for the spontaneous
breaking of
charge conjugation as discussed in ref.5).
Let us look first at the SU(2) case. Though the potential $A_0$ becomes
$-A_0^T$ under charge
conjugation (so in particular the colour diagonal matrix C will just
flip sign), this will not affect the quantum average 3.14(a), since it
is even in C. So the average of the trace of the loop is charge conjugation
even, and
no charge conjugation breaking effects can be measured with it.
In general, for any N, charge conjugation leaves $<P(A_0)>$ invariant as long
as it is
real.

In between different vacua however the order parameter takes on complex values,
(except for $SU(2)$ of course), but this lies outside the scope of
this paper.
We will come back to the problem of computing the constrained effective action
when the argument is a profile, as a function of one spatial
variable$^{16)}$.

In fig 4 and 5 the reader will find the effective one loop potential for
$SU(3)$, with
and without quarks. How the metastable state$^{15)}$ in fig 6, near
$C={2\pi\over
3}$, is sensitive to two loop contributions is shown in the same
figure.

The effects of lattice artifacts will be evaluated in ref 17). The Montecarlo
evaluation could be done eventually with the multicanonical ensemble$^{18)\
19)}$ ; this method avoids the difficulty of the peak between two vacua (see
fig
6), which bars the MC-access to this domain of $q$-values. Of course the
massless fermions remain a practical problem.

Some time ago$^{15}$, it was pointed out that the total  one loop effective
action, i.e. the sum of the free
energy at C=0, F(0), and $U^{(1)}(C)$ lead for high enough fermion
number $n_f$ and appropriate values of C in the q-valley to absurd
thermodynamical properties, in particular near the metastable minimum
in fig 6. If we are only interested in local minima the presence of
fermions is needed.

 But in a Montecarlo simulation one can apply an
external
field coupled to the Polyakov loop.

In this latter context
it is amusing to observe that the same happens
for a system with gluons alone. For example, the maximum in fig 3
correponding
to the center of the admitted values for SU(3), has negative entropy:
explicitely, the free energy for $N^2-1$ gluons$^{1)}$ is $-{1\over
{45}}{\pi}^2 T^4 (N^2-1)$. Adding to this the value of the one loop
result for the effective potential 2.23 in the maximum gives for SU(2)
${\pi^2\over{60}}T^4$ and for SU(3) ${8\pi^2\over {405}}T^4$,
producing a negative entropy.
This negative entropy may be a very serious default of the
perturbative
approach. Since at this order the free energy is just counting the degrees
of freedom, the extremist would say the negative entropy indicates that we have
overlooked
hitherto unknown degrees of freedom in QCD.
However, already the first derivative of the effective potential with
respect to the order parameter shows peculiar behaviour: instead of
the intuitively expected flat part- corresponding to the spontaneous
breaking of the symmetry- it has smooth behaviour. This smooth
behaviour is the opposite of what happens in Mean Field approximation,
namely overcooling, and it fits with the fact that
the second
derivative of the effective potential is indeed convex (again, in
contrast to Mean Field). Note that the
effective potential is only convex in terms of the trace of the loop,
not in terms of the background field C!
 Clearly some
important physics remains to be understood.

\b

\n{\bf ACKNOWLEDGEMENTS}

I'm indebted to Tanmoy Bhattacharya, Andreas Gocksch, Bernd Grossmann,
Frithjof Karsch, Keijo Kajantie, Robert Pisarski and Jay Watson for
useful correspondence and discussions, J. Ignatius for discussions and
providing me with some figures.

\vfill\eject

\n Appendix A. {\bf Gauge terms proportional to $(1-\xi)^2$ AND $(1-\xi)^3$}
\b
Gauge terms proportional to $(1-\xi)$ are treated in the main text, section
2.e.

Terms proportional to $(1-\xi)^2$ are involving contractions of the 3-vertices
with two momenta. This contraction simplifies considerably the vertex. Consider
diagram (a) in fig 1. Its contribution to the $(1-\xi)^2$ term equals :
$$\eqalignno{
{g^2\over
12}\int_{1,2}&\left\{{1\over l^4_1l^4_2l^2_3}(l^2_1(l_2\cdot l_3)^2-2(l_1\cdot
l_2)
(l_2\cdot l_3)(l_3\cdot l_1)+l^2_2(l_1\cdot l_3)^2)\right.\cr
&+(1\rightarrow 2,\ 2\rightarrow 3,\ 3\rightarrow 1)+
(1\rightarrow 3,\ 2\rightarrow 1,\ 3\rightarrow 2))&A.1\cr}$$
Because of the symmetry of the integration summation $\int_{1,2}$ over $l_1$
and $l_2$ and colour, the two terms in A.1, not explicitely written, give the
same result as the first one we will now further analyse.

We use momentum conservation :
$$l_1+l_2+l_3=0\eqno A.2$$
to write
$$2(l_i\cdot l_j)=l^2_k-l^2_i-l^2_j\eqno A.3$$
for any triple $(i,j,k)=(1,2,3)$.

The first term  $I$ in A.1 gives then :
$$I\equiv{g^2\over 48}\int_{1,2}\left({l^2_1\over l^4_2l^2_3}+
{l^2_3\over l^2_1l^4_2}+{1\over l^2_1l^2_3}-{2\over l^2_2l^2_3}
-{2\over l^4_2}+{2\over l^2_1l^2_2}\right)\eqno A.4$$
The last four terms add up to :
$${g^2\over 48}\widehat B^2\cdot\widehat B^2\eqno A.5$$
using
$$\int_{1,2}{1\over l^4_2}=0\eqno A.6$$
The first two terms give an equal contribution upon integration ; the second is
seen to be, using A.2 :
$${g^2\over 48}\int_{1,2}{\left(l^2_1+2l_1\cdot l_2+l^2_2\right)\over
l^2_1l^4_2}={g^2\over 48}\int_{1,2}\left({1\over l^4_2}+{2l_1\cdot l_2\over
l^2_1l^4_2}+{1\over l^2_1l^2_2}\right)\eqno A.7$$

The second term equals :
$${g^2\over 48}\int_{1,2}{2l_{10}l_{20}\over l^2_1l^4_2}\eqno A.8$$
because the correlation ${\vec l_{1}\cdot \vec l_{2}\over l^2_1l^4_2}$ is odd
in $\vec l_1$ (and $\bar l_2$).
Using A.4 to A.8 we get :
$$I={g^2\over 48}(4\widehat B_1\cdot\widehat B_3+3\widehat B_2\cdot\widehat
B_2)\eqno A.9$$
The dots mean summation over colour degrees of freedom :
$$\widehat
B_i\cdot\widehat
B_j\equiv\displaystyle\sum_{a,b,c}|f_{a,b,c}|^2B_i(C_a)B_j(C_b)$$

The third term in A.1 is through exchange $(1\rightarrow 2)$ identical to the
first one in A.4 we just computed.

The middle term in A.1 equals :
$$\eqalignno{
-{g^2\over
48}\int_{1,2}&(l^6_3+l^6_2+l^6_1-l^4_3l^2_1-l^4_3l^2_2-l^4_1l^2_2-l^4_1l^2_3\cr
&-l^4_2l^2_1-l^4_2l^2_3+2l^2_3l^2_2l^2_1){1\over l^4_1l^4_2l^2_3}&A.10\cr}$$
The first term in A.10 gives
$${-g^2\over 48}\int_{1,2}{l^4_3\over l^4_1l^4_2}\eqno A.11$$
whereas the other terms all cancel, using symmetry in the integration
variables,
and the result is for A.1 :
$${g^2\over 8}\left(4\widehat B_1\cdot\widehat B_3+3\widehat B_2\cdot\widehat
B_2-{1\over 2}\int_{1,2}{l^4_3\over l^4_1l^4_2}\right)\eqno A.12$$
The $(1-\xi)^2$ contribution from diagram $(b)$ in fig 1 can be worked out with
the same tricks and gives the same result as in A.12, but with opposite sign.
That is : the $(1-\xi)^2$ term from fig 1 is zero.

The $(1-\xi)^3$ term is zero, because the 3-vertex vanishes when
contracted with all three incoming momenta.

\vfill\eject

\n Appendix B. {\bf Quantum average of Polyakov loops}
\b
In this appendix the trace of the quantum average of the $n-th$ power of the
Polyakov-loop is worked out.

Let us introduce some notation. The matrix of order $g^2$ in eqn 3.1 is called
$L_2$ :
$$\eqalignno{
L_2\equiv
-g^2\int^{\beta}_0d\tau_1\int^{\tau_1}_0&d\tau_2e^{iC\tau_2}{1\over V}\int
d\vec x Q_0(\tau_2\vec x)e^{iC(\tau_1-\tau_2)}\cr
&Q_0(\tau_1\vec
x)e^{iC(\beta-\tau_1)}&B.1\cr}$$

Remember $C$ is a diagonal $N\times N$ matrix with $i^{th}$ entry $C_i$.

Now we have for the propagator in a finite volume from 2.21~:
$$\left<Q_0(p_0\vec p)Q_0(p_0'\vec p')\right>={1\over T}\delta_{p_0,-p_0'}
\delta_{\bar p,-\bar p'}{1\over V}\left({1\over p^2_0+\vec
p^2}-(1-\xi){p_0^2\over(p^2_0+\vec p^2)^4}\right)\eqno B.2$$
So :
$${1\over V}\int d\vec x\left<Q_0(\tau_1\vec x)Q_0(\tau_2\vec x)\right>=
T\sum_{p_0}\int{d\vec p\over (2\pi)^3}\left({1\over p^2_0+\vec
p^2}-(1-\xi){p_0^2\over(p^2_0+\vec
p^2)^4}\right)e^{ip_0(\tau_1-\tau_2)}\eqno B.3$$

in the infinite volume limit.

In calculating $Tr<L>$ we push $e^{iC\tau_2}$
through $Q^{ij}_0(\tau_2)\lambda^{ij}$ to the right, which causes an extra
phase
$e^{i(C_i-C_j)\tau_2}$ to $Q^{ij}_0(\tau_2)$. This phase adds to the Fourier
coefficent $e^{ip_0\tau_2}$ of $Q^{ij}_0(\tau_2)$ ; the same happens to the
Fourier coefficient of  $Q^{ij}_0(\tau_1)$, it becomes
$e^{i(p'_0-(C_i-C_j))\tau_2}$.

Having done this, integration over $\tau_2$ and $\tau_1$ as in B.1, gives eqn
3.9 in the text.

When computing the quantum average of $TrP^n(A),\ (n\geq 2)$ to order $g^2$, we
also need the first order matrix :
$$L_1\equiv ig\int^{\beta}_0d\tau e^{iC\tau}Q_0(\tau)e^{+iC(\beta-\tau)}\eqno
B.4$$
To order $g^2$ we get (using $P(A_0)=e^{iC\beta}+L_1+L_2)$ :
$$\eqalignno{
\left<TrP^n(A)\right>=&nTr\left<L_2\right>e^{i(n-1)C\beta}\cr
&+(n-1)Tr\left<L_1L_1\right>e^{i(n-2)C\beta}\cr
&+(n-2)Tr\left<L_1e^{iC\beta}L_1\right>e^{i(n-3)C\beta}\cr
&+\dots+Tr\left<L_1e^{i(n-2)C\beta}L_1\right>+0(g^4)&B.5\cr}$$
We now push in each term the matrices $e^{iC\tau_i}$ to the right, to have the
two potentials next to one another. This leads to extra phases as below eqn
B.1, and B.5 becomes, for a fixed  pairing $\left<Q^{ij}_0(\tau_2)
Q^{ji}_0(\tau_1)\right>\equiv\Delta^{ij}_{00}$~:
$$\eqalignno{
\left<TrP(A)^n\right>=&-g^2\left[\int^{\beta}_0d\tau_1\int^{\tau_1}_0d\tau_2
\left\{nTrD^{ii}\Delta_{00}^{ij}e^{ip^{ij}(\tau_2-\tau_1)}e^{inC\beta}\right.\right.
\cr
&\left.+nTrD^{jj}\Delta_{00}^{ij}e^{-ip^{ij}(\tau_2-\tau_1)}e^{inC\beta}\right\}\cr
&+\int^{\beta}_0d\tau_2\int^{\beta}_0d\tau_1
\left\{(n-1)Tr\left(D^{ii}\Delta_{00}^{ij}e^{ip^{ij}(\tau_2-\tau_1)iC_{ij}\beta}
e^{inC\beta}\right.\right.\cr
&+D^{jj}\Delta_{00}^{ij}e^{-ip^{ij}(\tau_2-\tau_1)iC_{ij}\beta}e^{inC\beta}\cr
&+(n-2)Tr\left(D^{ii}\Delta_{00}^{ij}e^{ip^{ij}(\tau_2-\tau_1)-2iC_{ij}\beta}
e^{inC\beta}\right.\cr
&\left.+D^{jj}\Delta_{00}^{ij}e^{-ip^{ij}(\tau_2-\tau_1)+2iC_{ij}\beta}e^{inC\beta}
\right)\cr
&+\dots\cr
&+Tr\left(D^{ii}\Delta_{00}^{ij}e^{ip^{ij}(\tau_2-\tau_1)-i(n-1)C_{ij}\beta}
e^{inC\beta}\right.\cr
&\left.+D^{jj}\Delta_{00}^{ij}e^{-ip^{ij}(\tau_2-\tau_1)+i(n-1)C_{ij}\beta}
e^{inC\beta}\right)&B.6\cr}$$

Do the $\tau$-integrals and obtain, using the notation in 3.10~:

$$\eqalignno{
{1\over
N}Tr\left<P^n(A_0)\right>=&{-g^2\over
N}\left[{\beta\over i}
nTr\left(D^{ii}-D^{jj}\right)\Delta_{1}^{ij}e^{inC\beta}\right.\cr
&+n\left(TrD^{ii}\Delta_{(2)}^{ij}e^{inC\beta}\left(1-e^{-iC_{ij}\beta}\right)
\right.\cr
&+\left.TrD^{jj}\Delta_{(2)}^{ij}e^{inC\beta}\left(1-e^{iC_{ij}\beta}\right)
\right)\cr
&+\Delta_{(2)}^{ij}|1-e^{iC_{ij}\beta}|^2\left\{
(n-1)Tr\left(D^{ii}e^{inC}e^{-iC_{ij}\beta}+D^{jj}e^{inC}e^{iC_{ij}\beta}
\right)\right.\cr
&+(n-2)Tr\left(D^{ii}e^{inC}e^{-i2C_{ij}\beta}+D^{jj}e^{inC}e^{i2C_{ij}\beta}
\right)\cr
&+\dots\cr
&\left.\left.+Tr\left(D^{ii}e^{inC}e^{-i(n-1)C_{ij}\beta}+D^{jj}e^{inC}e^{i(n-1)C_{ij}\beta}
\right\}\right)\vphantom{{\beta\over\tau}}\right]&B.7\cr}$$
The terms in the braces in B.7 do add up pairwise : the first and the last to
$n\ e^{i\left({n\over 2}-1\right)C_{ij}\beta}e^{i{n\over 2}(C_i+C_j)\beta}$,
the second and the one but last to
$n\ e^{-i\left({n\over 2}-1\right)C_{ij}\beta}\ e^{i{n\over
2}(C_i+C_j)\beta}$, etc... Thus gathering all terms proportional to
$\Delta_{(2)}(ij)$, we obtain as coefficient :
$$\eqalignno{
n\ &e^{i{n\over 2}(C_i+C_j)\beta}\left\{e^{i{n\over 2}C_{ij}\beta}
-e^{i\left({n\over
2}-1\right)C_{ij}\beta}+(2-e^{iC_{ij}\beta}-e^{-iC_{ij}\beta})\right.\cr
&\left.\left(
e^{i\left({n\over 2}-1\right)C_{ij}\beta}
+e^{i\left({n\over 2}-2\right)C_{ij}\beta}+e^{i\left({n\over
2}-3\right)C_{ij}\beta}+\dots\right)+c\cdot c\cdot\right\}&B.8\cr}$$
What is in the braces in B.8 adds up to zero. All what remains is the first
term
in B.7, which is nothing but eqn 3.15 in the main text.

 Amusingly, we can find another form of 3.17:
$${\underline{\delta}C_i-\underline{\delta}C_j\over {2\pi T}}={g^2N\over
(4\pi)^2}(2+1-\xi)B_1(C_{ij})\eqno 3.18$$
This form is only true in special directions in the Lie algebra, given by the generalized hypercharges: 
$$Y_k={1\over N} diag(k,k, ...k, k-N, k-N,....,k-N).$$
The entry k is integer, running from 1 to N-1. It is repeated N-k times, whereas k-N is repeated k times. In these directions the sign function has the wanted property $$\sum_l(\epsilon(C_{il})-\epsilon(C_{jl}))=N\epsilon(C_{ij}).$$

\vfill\eject

\n Appendix C {\bf Bernoulli polynomials}
\b
The Bernoulli polynomials come about naturally in thermal field theory. We
take from ref (14)
$$B_{2k}(x)\equiv\sum^{\infty}_{r=1}{1\over r^{2k}}(-)^{k-1}{2(2k)!
\over(2\pi)^{2k}}\cos r2\pi x\eqno C.1$$
and
$$2kB_{2k-1}(x)\equiv B'_{2k}(x).\eqno C.2$$

Explicitely one has on the interval $-1\leq x\leq 1$ :

$$\eqalignno{
&B_4(x)=x^2(1-|x|)^2-{1\over 30}\cr
&B_3(x)=x^3-{3\over 2}x^2\epsilon(x)+{1\over 2}x&C.3\cr
&B_2(x)=x^2-|x|+{1\over 6}\cr
&B_1(x)=x-{1\over 2}\epsilon(x).\cr}$$
$\epsilon(x)$ is the sign function $x/|x|$.

They are related to integrals of the type :
$$\widehat B_{d-2k}\equiv T\sum_{n_0}\int {d^{d-1}\bar
P\over(2\pi)^{d-1}}{1\over((2\pi Tn_0+2\pi Tx)^2+\bar P^2)^k}\eqno C.4$$
$$\widehat B_{d-2k+1}\equiv T\sum_{n_0}\int {d^{d-1}\bar
P\over(2\pi)^{d-1}}{(2\pi Tn_0+2\pi Tx)\over((2\pi Tn_0+2\pi Tx)^2+
\bar P^2)^k}\eqno C.5$$
$$\widehat B_d\equiv T\sum_{n_0}\int {d^{d-1}\bar
P\over(2\pi)^{d-1}}(\log ((2\pi Tn_0+2\pi Tx)^2+\bar P^2)-\log((2\pi
Tn_0)^2+\bar P^2))\eqno C.6$$
$d$ is the number of dimensions of space time. For $d=4$ we have
$$\widehat B_4(x)={2\pi^2\over 3}T^4\left(B_4(x)+{1\over 30}\right)\eqno C.7$$
$$\widehat B_3(x)={2\pi\over 3}T^3B_3(x)\eqno C.8$$
$$\widehat B_2(x)={T^2\over 2}B_2(x)\eqno C.9$$
$$\widehat B_1(x)=-{T\over 4\pi}B_1(x)\eqno C.10$$

Note that $\widehat B_4$ and $\widehat B_2$ are even in $x$, $\widehat B_3$
and $\widehat B_1$ odd. Hence $\widehat B_2$ and $\widehat B_1$ are not
analytic in $x=0$.

In deriving 4.3 from 4.7 and 4.8 we have used the identities :
$$B_1(x)B_3(x)=\widetilde B_4(x)+{1\over 4}\widetilde B_2(x)\eqno C.11$$
and
$$(B_2(x))^2=\widetilde B_4(x)+{1\over 3} \widetilde B_2(x)+B_2^2(0)\eqno C.12$$
and
$$\widetilde B_2(x)\equiv B_2(x)-{1\over 6}\eqno C.13$$

Since it occurs frequently we define
$$\tilde B_4\equiv B_4 +{1\over {30}}\eqno C.14$$

For the evaluation of $\Delta^{ij}_{(1)}$ in eqn 3.10 it is useful to do
the d-1 momentum integrations first, to find a $\zeta$-function of the
type$^{14)}$:
$$\zeta(z,q)=\sum_{n=0}^{\infty}{1\over {(n+q)^z}}\eqno C.15$$
with $z=4-d$.
Using$^{14)}$
$$\zeta{(-z,q)}=-{B^{\prime}_{z+2}(q)\over(z+1)(z+2)}=-{B_{z+1}(q)\over{z+1}}\eqno C.16$$
valid for z non-negative, one finds
$$T\sum_{n_0}\int {d\vec
p\over{{(2\pi)}^{d-1}}}{1\over{p_0^{ij}(p^{ij})^2}}= {1\over
{2\pi }}B_1(C_{ij}) \eqno C.17$$
as in 3.14(b).

The  gauge term in $\Delta^{ij}_{(1)}$ follows immediately from C.5
and C.10.

$\Delta^{ij}_{(2)}$ will have terms like in C.15, but now with z=5-d.
This gives us a pole$^{14}$ at d=4.
\vfill\eject

\n Appendix D {\bf Some group theory relations and two relations for the gluon
potential}
\b

We introduce two unitarily related bases on the Lie algebra of $SU(N)$.

\item{1)} The Gell Mann basis $\lambda_a$, which is hermitean, traceless and
normalised
$$Tr\lambda_a\lambda_b={1\over 2}\delta_{ab}\eqno D.1$$
\m
\item{2)} The Cartan basis $\lambda^{ij},\ \lambda^d\quad(i\not = j),\quad
d=2,3,\dots , N$, with :
$$(\lambda^{ij})_{kl}\equiv{1\over\sqrt{2}}\delta_{ik}\delta_{jl}\ ,\
\lambda_d={1\over r_d}diag(1,1,\dots,1-d,0,0,0),\ r_d\equiv\sqrt{2d(d-1)}.$$

In both cases we define the structure constants $f$ as~:
$$if_{abc}\equiv 2TrM_a[M_b,M_c]\eqno D.2$$
In basis 2) we have
$$|f_{ij,jk,ki}|^2={1\over 2}\eqno D.3$$
and
$$\sum^N_{d=2}|f_{ij,ji,d}|^2=1\eqno D.4$$

All other structure constants vanish in basis 2).

It is useful to note that there are $2N(N-1)(N-2)$ $f$'s like in D.3, and that
there are $3N(N-1)$ sums like in D.4.

So, in basis 1) we have
$$\sum_{a,b,c}|f_{a,b,c}|^2=N(N^2-1)\eqno D.5$$
and in basis 2) we find again :
$$\sum_{a,b,c}|f_{a,b,c}|^2={1\over 2}\cdot 2N(N-1)(N-2)+3N(N-1)=N(N^2-1)\eqno
D.6$$
as it should, because of the unitary relationship between the two sets.

For the fermionic two-loop contribution 5.4 we need another identity for the
coefficients $r_k$ defined under D.1 :
$${(d-1)^2\over
r^2_d}+\sum^N_{k=d+1}{1\over r^2_k}={1\over 2}{N-1\over N}\eqno D.7$$
D.7 is true for all $d=2,\dots, N-1$. It is useful when computing the
exchange of all diagonal gluons in fig 1(d).

Let us finally relate the determinant $\vert\vert
t^{(1)}_{k,d}(C)\vert\vert$
(appearing in the first term of eqn 2.21) to the van der Monde
determinant
$$\prod_{1\leq i\le j\leq N}(\exp{iC_i}-\exp{iC_j})\eqno D.8$$
This equality is true up to a factor independent of C.
This determinant was defined as the Jacobian between the variables
$t_1(C),t_2(C),....t_{N-1}(C)$, eqn 2.1(a), and the variables
$C_d\equiv Tr\lambda_d C$, d=2,....N. From this it follows
immediately, that a generic matrix element reads:
$$
t^{(1)}_{k,d}(C)={k\over{Nr_d}}(\exp{ikC_1}+\exp{ikC_2}+...
-(d-1)\exp{ikC_d})\eqno D.9$$
{}From this expression it is easy to see that the determinant of the
(N-1)x(N-1)
matrix $t^{(1)}_{k,d}(C)$ can be written as the determinant of the NxN van der
Monde
matrix, by adding and subtracting columns, and D.9 follows.
\vfill\eject

\noindent{\bf REFERENCES}
\medskip
\parindent=9truemm

\item{\hbox to\parindent{\enskip 1)}\hfill}
D. Gross, R. Pisarski, L. Yaffe, Rev. Mod. Phys. {\bf 53} (1981) 43

\item{\hbox to\parindent{\enskip }\hfill}N. Weiss, Phys. Rev. D. {\bf 24}
(1981) 75 ; D {\bf 25} (1982) 2667

\item{\hbox to\parindent{\enskip 2)}\hfill}S. Huang, Y. Potuin, C.
Rebbi, and S. Sanielevici, Phys. Rev. D {\bf 42} (1990) 2864

\item{\hbox to\parindent{\enskip 3)}\hfill}K. Kajantie, L. K\" arkainen, K.
Rummukainen, Nucl. Phys. B {\bf 357} (1991) 693

\item{\hbox to\parindent{\enskip 4)}\hfill}T. Bhattacharya, A. Gocksch, C.P.
Korthals Altes,  R.D. Pisarski, Phys. Rev. Lett. {\bf 66} (1991) 998

\item{\hbox to\parindent{\enskip 5)}\hfill}T. Bhattacharya, A. Gocksch, R.D.
Pisarski, Nucl. Phys. B {\bf 383}, (1992) 497

\item{\hbox to\parindent{\enskip 6)}\hfill}K. Enkvist, K. Kajantie, Z. Phys. C
{\bf 47} (1990) 291

\item{\hbox to\parindent{\enskip 7)}\hfill}R. Anishetty, J. Phys. G. {\bf 10}
(1984) 439

\item{\hbox to\parindent{\enskip 8)}\hfill}R. Fukuda, E. Kyriakopoulos, Nucl.
Phys. B {\bf 85} (1975) 354

\item{\hbox to\parindent{\enskip 9)}\hfill}L. O'Reifearteigh, A. Wipf, H.
Yoneyama, Nucl. Phys. B {\bf 271} (1986) 653

\item{\hbox to\parindent{\enskip 10)}\hfill}C. Becchi, R. Rouet, R. Stora, Ann.
Phys. (N.Y.) {\bf 98} (1976) 287

\item{\hbox to\parindent{\enskip 11)}\hfill}C.P. Korthals Altes, in Progress in
Gauge Theory, Carg\`ese 1983, eds 't Hooft et al, Plenum 1984.

\item{\hbox to\parindent{\enskip 12)}\hfill}V.M. Belyaev, Phys. Lett. B {254}
(1991) 153

\item{\hbox to\parindent{\enskip 13)}\hfill}A. Gocksch, R.D. Pisarski,
BNL-GP-1/93

\item{\hbox to\parindent{\enskip 14)}\hfill}I.M. Ryshik, I.S. Gradsteijn, Table
of Integrals Series and Products, 9.622, Academeic Press (1965)

\item{\hbox to\parindent{\enskip 15)}\hfill}V. Dixit, M. Ogilvie, Phys. Lett. B
{\bf 269}, 353 (1991)

\item{\hbox to\parindent{\enskip 16)}\hfill}To be published.

\item{\hbox to\parindent{\enskip 17)}\hfill}T. Bhattacharya, C.P.
Korthals Altes, in preparation.

\item{\hbox to\parindent{\enskip 18)}\hfill}B. Berg, T. Neuhaus, Phys. Rev.
Lett. {\bf 68} (1992) 9

\item{\hbox to\parindent{\enskip 19)}\hfill}B. Grossmann, M.L. Laursen, in
''Dynamics of First Order Phase Transitions'', eds H. J. Herrmann, W. Janke,
F. Karsch, World Scientific, 1992

\item{\hbox to\parindent{\enskip 20)}\hfill}C.P.Korthals Altes, in Hot
Summer Daze, Proceedings of the BNL Summer Study on QCD at non-zero
temperature and density, Eds A.Gocksch and R.Pisarski, World
Scientific, 1992
\vfill\eject
\nopagenumbers

{\parindent=1cm\narrower
\centerline{\includegraphics[width=10cm]{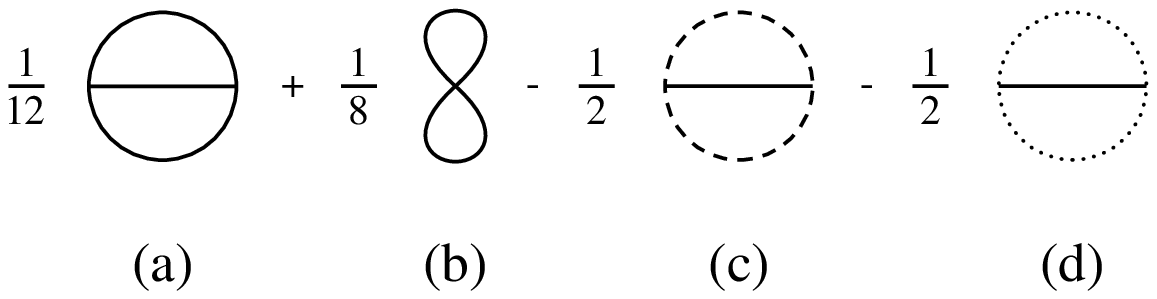}}
\noindent Fig 1: Graphs obtained from 2.12 with the topology of the free
energy(a,b,c). The two loop fermion graph is shown in 1.d. The symmetry
factor is explicitly written. Continous (dashed) lines are gluons (ghosts).
Constant
background gauge is used, hence all energies are shifted through a
constant amount. Their contribution  $U^{(2)}_f$ is given in 4.7 (for $\xi=1$) and
2.35 (if $\xi\neq 1$) for the gluons, and in 5.4 for the fermions.\par}

\vfill\eject

{\parindent=8truemm\narrower
\centerline{\includegraphics[width=10cm]{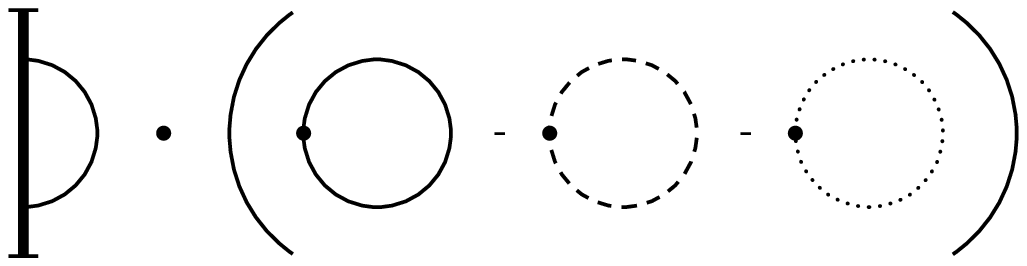}}
\noindent Fig 2: Zero momentum insertion accompanied by renormalisation of the
Polyakov-loop, $U^{(2)}_p$ in eqn.2.28.The dot on the loops are the
zero-momentum insertions 2.28(b).\par}

\vfill\eject

{\parindent=10truemm\narrower
\centerline{\includegraphics[width=10cm]{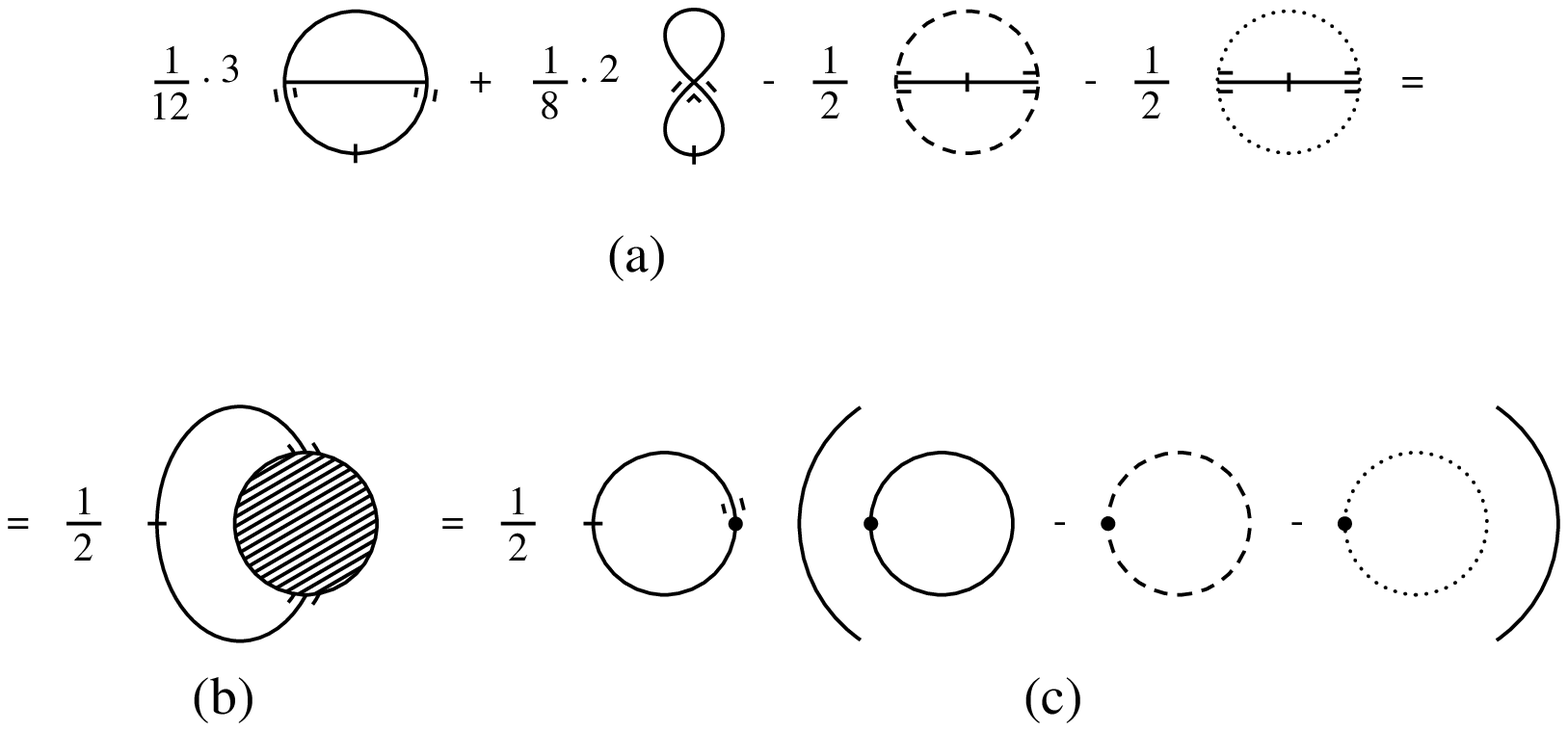}}
\noindent Fig 3: Relation between gauge variation of the free 
energy graphs (a),
the gauge variation of the gluon self energy (b), and the zero-momentum
insertion into the (cross-hatched) one loop free energy (c) through the
BRST identity 2.32. The double bars represent the contraction of the
momentum of the line into the vertex. The loop in the left part of
fig.3(c) equals the gauge variation of the renormalisation of the Polyakov
loop in fig 2.\par}

\vfill\eject

{\parindent=10truemm\narrower
\centerline{\includegraphics[width=10cm]{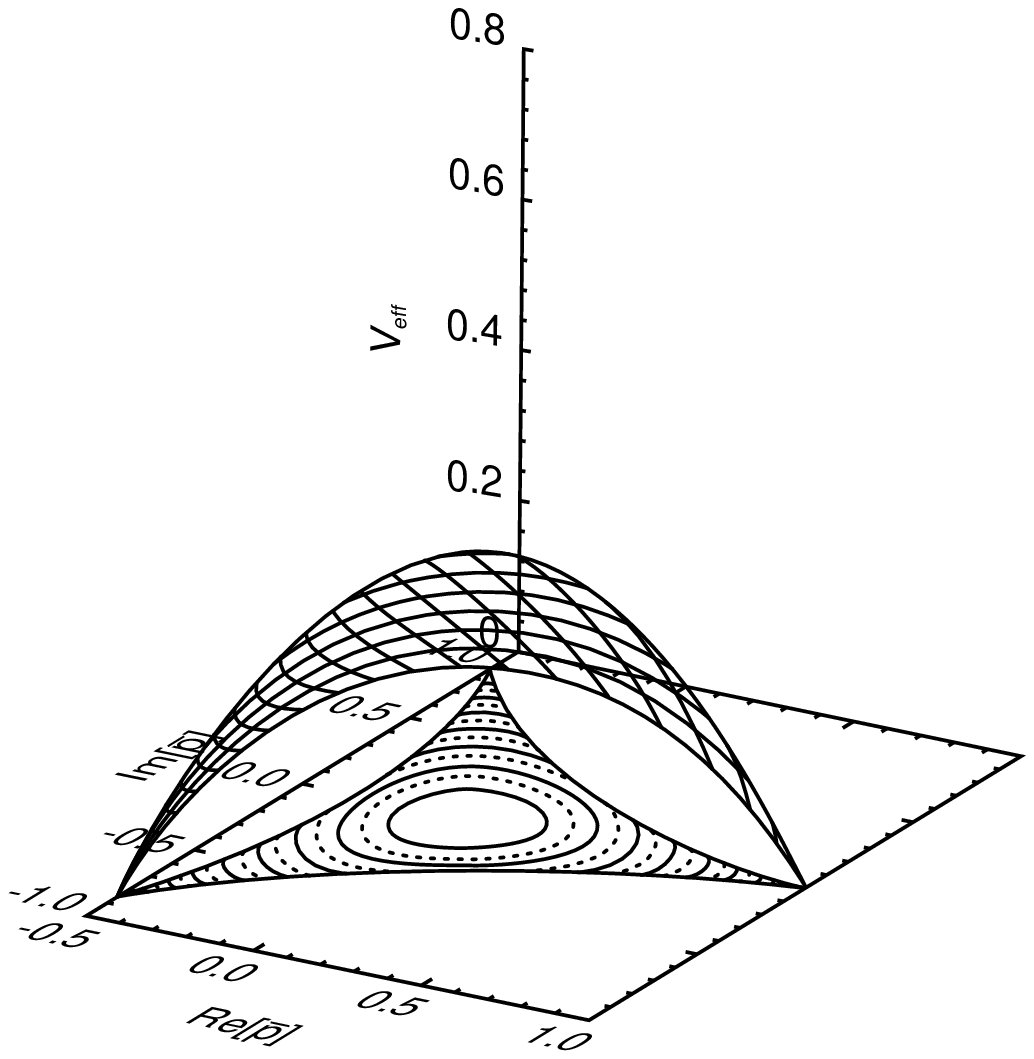}}
\noindent Fig 4: The potential 4.2 and 4.3 for $N=3$ and $n_f=0$, as function of
real and imaginary part of
the loop. $V_{eff}$ stands for ${3\over {8\pi^2 T^4}}U$. \par}

\vfill\eject

{\parindent=10truemm\narrower
\centerline{\includegraphics[width=10cm]{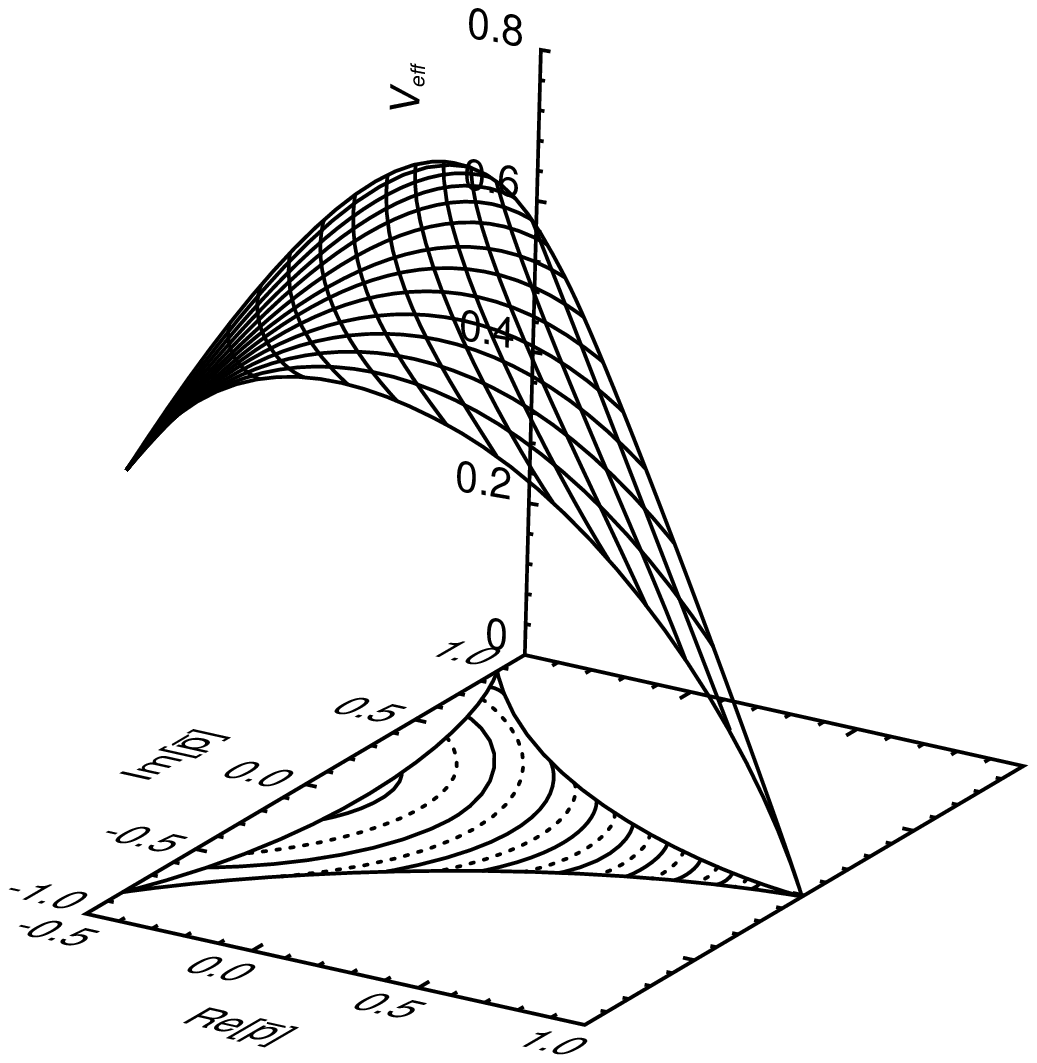}}
\centerline{Fig 5: As in Fig 4, with $n_f=2$, using 5.2 and 5.4.}\par}

\vfill\eject

{\parindent=10truemm\narrower
\centerline{\includegraphics[width=10cm]{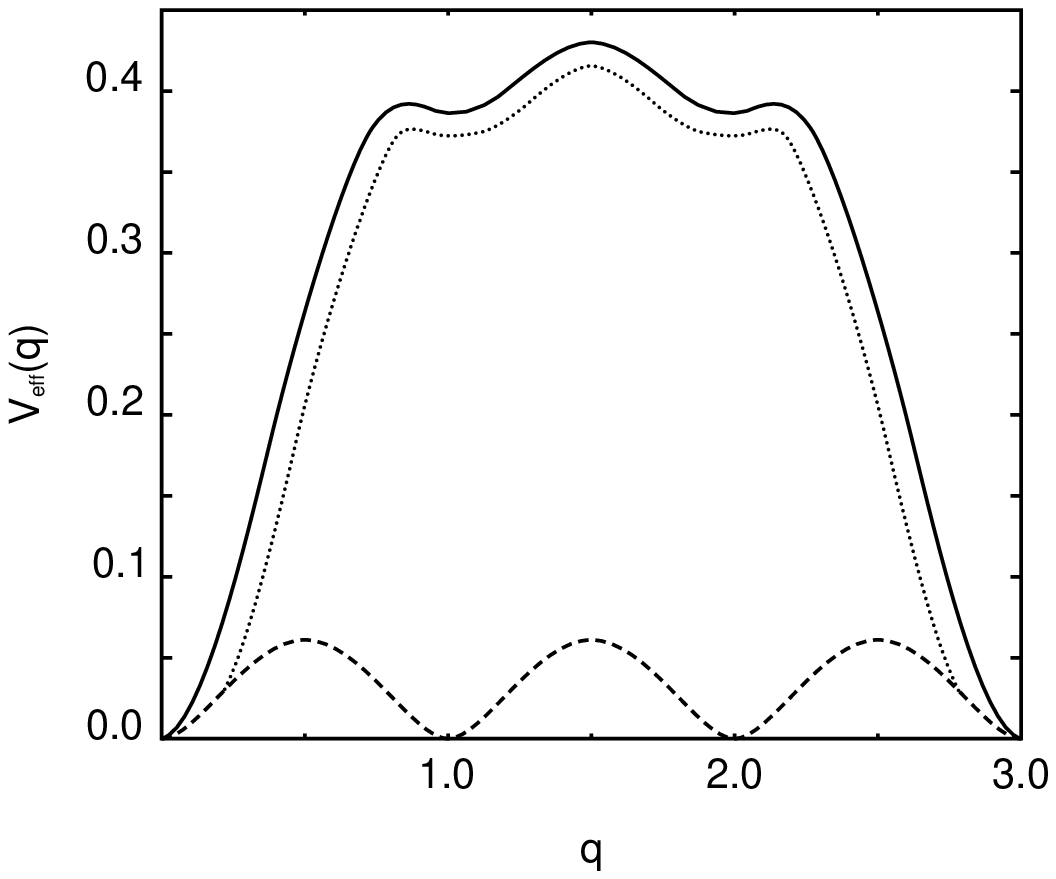}}
\noindent Fig 6: The q-valley profile for the case N=3 and $n_f=6$. Solid
curve is the one loop result 4.2 and 5.2., the dotted curve the one
and two loop result at $\alpha_s$=0.1.The dashed curve is the pure
gluon result. \par}

\end